\documentclass[prd,twocolumn,eqsecnum,amsfonts,amssymb]{revtex4}

\usepackage{graphicx}
\usepackage{subfig}

\usepackage{bm}

\newcommand{\beq}{\begin{equation}}
\newcommand{\eeq}{\end{equation}}
\newcommand{\beqs}{\begin{eqnarray}}\newcommand{\eeqs}{\end{eqnarray}}
\newcommand{\lsim}{\mathrel{\raisebox{-
.6ex}{$\stackrel{\textstyle<}{\sim}$}}}
\newcommand{\gsim}{\mathrel{\raisebox{-
.6ex}{$\stackrel{\textstyle>}{\sim}$}}}

\begin{document}

\title{Using $\bar p p$ and $e^+e^-$ Annihilation Data to Refine Bounds on 
the Baryon-Number-Violating Dinucleon Decays $nn \to e^+e^-$ and
$nn \to \mu^+\mu^-$}

\author{Shmuel Nussinov$^a$ and Robert Shrock$^b$}

\affiliation{(a) \ School of Physics and Astronomy, Tel Aviv University, 
6997801 Tel Aviv, Israel} 

\affiliation{(b) \ C. N. Yang Institute for Theoretical Physics and
Department of Physics and Astronomy, \\
Stony Brook University, Stony Brook, New York 11794, USA }
\begin{abstract}

  We use $\bar p p$ and $e^+e^-$ annihilation data to further
  strengthen lower bounds on the partial lifetimes for the
  baryon-number-violating dinucleon decays $nn \to e^+ e^-$ and
  $nn \to \mu^+\mu^-$. 

\end{abstract}

\maketitle


\section{Introduction}
\label{intro_section}

In Ref. \cite{dnd}, lower limits on the partial lifetimes $\tau/BR
\equiv \Gamma^{-1}$ for a number of $\Delta B=-2$, $\Delta L=0$
dinucleon decays were presented, including $nn \to e^+e^-$, $nn \to
\mu^+\mu^-$, $nn \to \nu_\ell \bar\nu_\ell$, and $np \to
\ell^+\nu_\ell$, where $\ell=e, \ \mu, \ \tau$.  (Here, for the decay
of an initial state to a given final state, $\Gamma$ and $BR$ denote
the decay rate and branching ratio, and $\tau$ denotes the mean life
of the initial state.)  The lower bounds obtained in \cite{dnd} were
substantially stronger than limits from direct experimental
searches. In this paper we use data on $\bar p p$ and $e^+e^-$
annihilation to further improve the lower limits on the partial
lifetimes for $nn \to e^+e^-$ and $nn \to \mu^+\mu^-$ decays.

The violation of baryon number, $B$, is expected to occur in nature,
because this is one of the necessary conditions for generating the
observed baryon asymmetry in the universe \cite{sakharov}.  Baryon
number violation (BNV) is, indeed, predicted in many ultraviolet
extensions of the Standard Model (SM), such as grand unified
theories. A number of dedicated experiments have been carried out
since the early 1980s to search for proton decay (and the decay of
neutrons bound in nuclei). These experiments have obtained null
results and have set stringent lower limits on the
partial lifetimes for such $\Delta B=-1$ baryon-number-violating nucleon
decays. A particularly strong lower bound, $\tau/BR > 1.6 \times 10^{34}$ yrs, 
has been set by the Super-Kamiokande (SK) experiment 
for the decay channel $p \to e^+ \pi^0$ \cite{abe17},
which can be clearly identified in the
water Cherenkov detector of this experiment.
(This and other experimental limits are quoted at the 90
\% confidence level, CL.)

A different type of baryon number violation has also received
attention, namely $n - \bar n$ oscillations, which have $|\Delta B|=2$
\cite{kuzmin}-\cite{sk_nnbar}. It was observed early on that $n-\bar
n$ oscillations might provide the source of baryon number violation
necessary for baryogenesis \cite{kuzmin}.  We denote the $n-\bar n$
transition amplitude as $\langle \bar n | {\cal H}_{eff}| n\rangle
\equiv \delta m$.  In (field-free) vacuum, the Hamiltonian matrix has
diagonal elements $\langle n | {\cal H}_{eff} | n \rangle = \langle
\bar n | {\cal H}_{eff} |\bar n \rangle = m_n - i(\lambda_n/2)$, where
$\lambda_n = 1/\tau_n$ is the decay rate of a free neutron. The
diagonalization of this matrix yields the mass eigenstates $|n_\pm
\rangle =( |n \rangle \pm |\bar n \rangle )/\sqrt{2}$, with
eigenvalues $m_{\pm} = (m_n \pm \delta m) - i\lambda_n/2$.  Starting
with a pure $|n\rangle$ state at $t=0$, there is then a probability
for this to be a $|\bar n\rangle$ at time $t > 0$ given by $|\langle
\bar n|n(t) \rangle|^2 = [\sin^2(t/\tau_{n \bar n})] e^{-\lambda_n
  t}$, where $\tau_{n \bar n} = 1/|\delta m|$.  An experiment at the
Institut Laue-Langevin searched for $n-\bar n$ oscillations using a
neutron beam from a reactor and obtained the lower bound $\tau_{n \bar
  n} > 0.86 \times 10^8$ sec, i.e., $|\delta m| < 0.77 \times
10^{-29}$ MeV \cite{ill}.

The presence of a nonzero transition amplitude $\langle \bar n | {\cal
  H}_{eff} | n \rangle$ means that a physical neutron state
$|n\rangle_{\rm phys.} = \cos \theta_m |n\rangle + \sin\theta_m |\bar
n \rangle$ in a nucleus has an admixture of $|\bar n\rangle$. This
admixture has a very small coefficient,
\beq
\sin\theta_m \simeq \theta_m
\sim \frac{|\delta m|}{[(V_{n,R}-V_{\bar n,R})^2+V_{\bar n, I}^2]^{1/2}} \lsim
10^{-31} \ ,
\label{thetam}
\eeq
where $V_n = V_{n,R}$ and $V_{\bar n} = V_{\bar n,R}+i V_{\bar n,I}$
denote the potentials of the $n$ and $\bar n$ in the nucleus.
As reflected by the imaginary term $iV_{\bar n,I}$ in $V_{\bar n}$, the
small admixture of $|\bar n\rangle$ in $|n\rangle_{phys.}$ leads
to annihilation with a neighboring neutron or proton in the
nucleus, and thus to $\Delta B = -2$ dinucleon decays. Owing to the
dominance of strong over electroweak interactions, these dinucleon
decays yield mainly hadronic final states, typically comprised of
multiple pions.  The small coefficient $\theta_m$ is compensated by
the large number $ \sim 10^{33}$ of nucleons in a nucleon
decay detector, so nucleon decay experiments are also sensitive to
these $\Delta B =-2$ dinucleon decays (a recent review is
\cite{nnbar_white}).

Because the operators that contribute to baryon-number-violating
decays of individual nucleons are four-fermion operators with
coefficients of mass dimension $-2$, while the operators that
contribute to $n-\bar n$ transitions and the associated dinucleon
decays are six-quark operators with coefficients of mass dimension
$-5$, it follows that, if the physics responsible for baryon number
violation were characterized by a single mass scale, $M_{BNV}$, then nucleon
decays would be much more important than $n-\bar n$ oscillations as a
manifestation of baryon number violation.  However, there are examples
of beyond-Standard-Model (BSM) physics in which BNV nucleon decay is
absent \cite{mm80} or is suppressed well below observable levels
\cite{nnb02}, so that $n-\bar n$ oscillations and the associated
$\Delta B =-2$ dinucleon decays are the main manifestation of baryon
number violation and can occur at levels comparable to current
bounds. Some further studies of such models include
\cite{wise,bvd,nnblrs}.

There is thus strong motivation to investigate the implications of
current experimental limits on $\Delta B=-2$ dinucleon decays.  Using
a minimal effective field theory approach, Ref. \cite{dnd} derived
approximate relations between the rates for dinucleon decays to
hadronic final states and to various $\Delta L=0$ dilepton final
states and combined these with experimental lower bounds on the
partial lifetimes for these hadronic dinucleon decays to infer rough
lower bounds on the dinucleon decays to dileptons.  In the present
work we shall use $\bar p p$ and $e^+e^-$ annihilation data to
strengthen the lower bounds obtained in Ref. \cite{dnd} on the partial
lifetimes for the dinucleon decays $nn \to \ell^+\ell^-$, where $\ell$
denotes $e$ or $\mu$.


\section{Background}
\label{background_section}

We first recall some relevant background. In the presence of a nonzero
$n-\bar n$ transition amplitude $\delta m$ and the associated
dinucleon decays, the rate for matter instability is
\beq
\Gamma_{m.i.} \equiv \frac{1}{\tau_{m.i.}} \simeq
\frac{2(\delta m)^2 |V_{\bar n I}|}
     {(V_{n R} - V_{\bar n R})^2 + V_{\bar n I}^2} \ .
     \label{gamma_mi}
\eeq
It follows that $\tau_{m.i.} \propto (\delta m)^{-2} = \tau_{n \bar
  n}^2$.  Explicitly, $\tau_{m.i.} = R \, \tau_{n \bar n}^2$, where
the factor $R \sim O(10^2) \ {\rm MeV} \simeq 10^{23}$ sec$^{-1}$
depends on the nucleus.  The SK experiment has set the best limit this
type of matter instability \cite{sk_nnbar}, $\tau_{m.i.} > 1.9 \times
10^{32}$ yr.  Antiproton annihilation on hydrogen yields multipion
final states with average multiplicities of $\sim 5$
\cite{amsler91,amsler98}. Monte Carlo simulations that account for
the absorption of $\bar n$ annihilation pions on their way out of the
${}^{16}$O nucleus have been carried out in Ref. \cite{sk_nnbar}.
These simiulations yield considerably lower average pion
multiplicities, namely 3.5 and 2.2 for total and charged pion
multiplicities resulting from a an $\bar n$ annihilation in a
${}^{16}$O nucleus \cite{sk_nnbar}.  Consequently, there is a
substantially larger probability for two-pion final states to occur in
antinucleon-nucleon annihilation in the ${}^{16}$O nuclei in the SK
detector than in $\bar p p$ annihilation. The most restrictive lower
bound on the partial lifetime of an exclusive $nn$ dinucleon decay is
for di-neutrons in ${}^{16}$O \cite{sk_dinucleon_to_pions}, namely 
\beq
\Gamma^{-1}_{nn \to 2\pi^0} > 4.04 \times 10^{32} \ {\rm yr} \ . 
\label{tau_nn_to_2pi0_sk}
\eeq

The leading contribution to the decay $nn \to \ell^+ \ell^-$ is
described by a Feynman diagram in which the $|\bar n\rangle$ component
in an initial $|n\rangle_{\rm phys.}$ annihilates with a neighboring
$n$, producing a virtual photon $\gamma$ in the $s$-channel, which
then materializes into the final-state $\ell^+ \ell^-$ pair. There is
also a weak neutral-current contribution from a diagram with a virtual
$Z$ boson in the $s$-channel, but this is heavily suppressed by the
factor $(2m_N)^2/m_Z^2 < 10^{-3}$.  Let us denote the four-momentum of
the virtual photon as $q$ and the four-momenta of the $\ell^-$
and $\ell^+$ as $p_2$ and $p_1$, with $q=p_1+p_2$ and
$q^2=s=(2m_N)^2$. Neglecting the heavily suppressed weak neutral-current
contribution, and neglecting small effects due to Fermi motion, 
the amplitude for $nn \to \ell^+\ell^-$ is 
\beq
A_{nn \to \ell^+ \ell^-} =
(\delta m) \, e^2 \, \langle 0 | J_{em}^\lambda |n \bar n\rangle \,
\frac{1}{q^2} \, [\bar u(p_2) \gamma_\lambda v(p_1)] \ , 
\label{amp_nn_to_ellellbar_photon}
\eeq
where $\delta m$ represents the initial $n-\bar n$ transition
amplitude, and $e=\sqrt{4\pi \alpha_{em}}$ and $J_{em}^\lambda$ denote the
electromagnetic coupling and current. 

It follows that
\beqs
\Gamma_{nn \to \ell^+\ell^-} & \sim &
P \, e^4 \,
\frac{R_2^{(\ell^+\ell^-)}}
     {R_2^{(2\pi^0      )}}    \, \Gamma_{nn \to 2\pi^0} \cr\cr
&\sim& P \, e^4 \, \Gamma_{nn \to 2\pi^0} \ ,
\label{gamma_nn_to_ellellbar}
\eeqs
where $P$ denotes the probability that the total angular momentum of
the initial $nn$ state is greater than 0 and the initial state has the
appropriate quantum numbers to produce a nonzero amplitude $A_{nn \to
  \ell^+\ell^-}$. Note that a $J=0$ initial $nn$ state yields a
vanishing coupling $\propto q_\lambda [\bar v(p_2)\gamma^\lambda
  u(p_1)]=0$ with the lepton electromagnetic current bilinear. This
estimate made use of the fact that the ratio of two-body phase space
factors $R_2^{(\ell^+\ell^-)}/R_2^{(2\pi^0)}$ is very close to unity
for both $\ell=e$ and $\ell=\mu$.  Combining
(\ref{gamma_nn_to_ellellbar}) with the experimental lower limit
(\ref{tau_nn_to_2pi0_sk}) for a di-neutron in an ${}^{16}$O nucleus,
Ref. \cite{dnd} then obtained the rough estimate for the lower bound
on the partial lifetime (i.e., inverse decay rate $\Gamma^{-1}$) for
$nn \to \ell^+\ell^-$ in an ${}^{16}$O nucleus:
\beqs
\Gamma^{-1}_{nn \to \ell^+\ell^-} && \gsim
P^{-1} \, (5 \times 10^{34} \ {\rm yr})
\cr\cr
&& \gsim 5 \times 10^{34} \ {\rm yr} \ {\rm for} \ \ell=e, \ \mu \ .
\label{tau_limit_nn_to_ellellbar}
\eeqs
%


\section{Application of $\bar p p$ and $e^+e^-$ Annihilation Data}
\label{pbarp_section}

We next improve the rough lower limit (\ref{tau_limit_nn_to_ellellbar}) in \cite{dnd}
by using $\bar p p$ and $e^+e^-$ annihilation data.  For a given reaction or decay,
let $s_i$ denote an initial state and let $s_a$ and $s_b$ denote two
(kinematically allowed) final states.
It will be convenient to introduce the compact notation 
\beq
R^{(s_i)}_{s_a/s_b} \equiv \frac{\Gamma_{s_i \to s_a}}{\Gamma_{s_i \to s_b}} 
= \frac{BR(s_i \to s_a)}{BR(s_i \to s_b)} \ .
\label{ratio_gen}
\eeq
We will calculate $R^{(\bar n n)}_{\ell^+\ell^-/2\pi^0}$ as an input for
$R^{(n n)}_{\ell^+\ell^-/2\pi^0}$.  Our input data will be from experiments
on $\bar p$ annihilation.  Therefore, it will be useful to reexpress the ratio
$R^{(\bar n n)}_{\ell^+\ell^-/2\pi^0}$ in terms of the ratio
$R^{(\bar p p)}_{\ell^+\ell^-/2\pi^0}$ multiplied by appropriate factors. Thus, for 
$\ell = e, \ \mu$, we write 
\begin{widetext}
\beq
R^{(\bar n n)}_{\ell^+\ell^-/2\pi^0} \equiv 
\frac{\Gamma_{\bar n n \to \ell^+\ell^-}}
     {\Gamma_{\bar n n \to 2\pi^0      }}
     = \Bigg [ \frac{ \frac{\Gamma_{\bar n n \to \ell^+\ell^-}}
                           {\Gamma_{\bar p p \to \ell^+\ell^-}} }
                    { \frac{\Gamma_{\bar n n \to 2\pi^0} }{\Gamma_{\bar p p \to 2\pi^0}} } \Bigg ] \,
\frac{\Gamma_{\bar p p \to \ell^+\ell^-} }{ \Gamma_{\bar p p \to 2\pi^0} } = 
 \Bigg [ \frac{ \frac{\Gamma_{\bar n n \to \ell^+\ell^-}}
                           {\Gamma_{\bar p p \to \ell^+\ell^-}} }
                    { \frac{\Gamma_{\bar n n \to 2\pi^0} }{\Gamma_{\bar p p \to 2\pi^0}} } \Bigg ] \,
\frac{BR(\bar p p \to \ell^+\ell^-) }{ BR(\bar p p \to 2\pi^0) } \ .
\label{nbn_ee_over_nbn_2pi0}
\eeq
\end{widetext}
From the isospin invariance of strong interactions, it follows that
\beq
\frac{\Gamma_{\bar n n \to 2\pi^0} }{\Gamma_{\bar p p \to 2\pi^0} } = 1 \ , 
\eeq
up to small corrections such as those due to electromagnetism.

Next, we focus on the case $\ell=e$ and make use of experimentally
measured quantities.  Since photon exchange in the $s$ channel makes
by far the dominant contribution to the reactions $\bar n n \to
e^+e^-$ and $\bar p p \to e^+e^-$ and since electromagnetic reactions
are invariant under time reversal, we will use experimental data on
the reactions $e^+e^- \to \bar p p$ and $e^+e^- \to \bar n n$ to
determine the ratio $\Gamma_{\bar n n \to e^+e^-}/\Gamma_{\bar p p \to
  e^+e^-}$ in the $\ell=e$ special case of
Eq. (\ref{nbn_ee_over_nbn_2pi0}).  The $e^+e^- \to \bar p p$ cross
section at center-of-mass energies $\sqrt{s}$ near threshold has been
measured in a number of experiments, e.g., at Orsay
\cite{orsay_pbarp}, Frascati \cite{fenice_pbarp}, BEPC
\cite{bes2_pbarp}, SLAC \cite{babar_pbarp}, and Novosibirsk
\cite{vepp2015,vepp2016}. For $\sqrt{s}$ beyond the kinematic zero at
threshold, this cross section is relatively flat in the interval $I:
\quad 1.9 < \sqrt{s} \lsim 2.0$ GeV, with the value
\beq
\sigma(e^+e^- \to \bar p p) \simeq 0.9 \pm 0.1 \ {\rm nb} .
\label{sigma_ee_to_pbarp_exp}
\eeq
The cross section $\sigma(e^+e^- \to \bar n n)$ was measured in an
early experiment by the FENICE Collaboration at ADONE
\cite{fenice_nbarn}, and more recently in experiments at Novosibirsk,
with the result \cite{achasov2014,vepp2015,vepp2016}
\beq
\sigma(e^+e^- \to \bar n n) \simeq 0.85 \pm 0.20 \ {\rm nb} 
\label{sigma_ee_to_nbn_exp}
\eeq
for $\sqrt{s} \in I$. The uncertainties listed here 
are estimates based on the comparison of values measured at a given 
$\sqrt{s}$ by the different experiments, as weighted by their error
bars. In passing, it is interesting to note that the $e^+e^- \to \bar
p p$ and $e^+e^- \to \bar n n$ cross sections in this energy interval
are nearly equal, to within experimental uncertainties, despite the
fact that the proton is charged while the neutron is neutral.  (A
review of results on $e^+e^- \to \bar p p$ and $e^+e^- \to \bar n n$
up to 2013 is given in \cite{denig_salme}.)  Using time reversal invariance, we
thus obtain
\beq
\frac{\Gamma_{\bar n n \to e^+e^-}}{\Gamma_{\bar p p \to e^+e^-}} \simeq  
\frac{\sigma_{e^+e^- \to \bar n n; I}}{\sigma_{e^+e^- \to \bar p p; I}} \simeq 0.9\ , 
\label{nn_ee_over_pp_ee}
\eeq
where the subscript $I$ indicates that the cross sections on the
right-hand side of (\ref{nn_ee_over_pp_ee}) were measured in the
interval $\sqrt{s} \in I$ near threshold, but beyond the kinematic
falloff at threshold.

Finally, we need to determine the third ratio in the $\ell=e$ special case of
Eq. (\ref{nbn_ee_over_nbn_2pi0}), $BR(\bar p p \to e^+ e^-)/BR(\bar p p \to 2\pi^0)$.
Measurements of the numerator of this ratio with stopped antiprotons include a CERN
experiment that obtained $BR(\bar p p \to e^+e^-)=(3.2 \pm 0.9) \times
10^{-7}$ \cite{cern_pbp_ee} and the subsequent PS170 experiment at
LEAR (Low Energy Antiproton Annihilation Ring) at CERN, which obtained
the more accurate value \cite{ps170}
\beq
BR(\bar p p \to e^+e^-)=(3.58 \pm 0.10) \times 10^{-7} \ .
\label{br_pbp_to_ee_exp}
\eeq

Several experiments have measured $BR(\bar p p \to 2\pi^0)$ for $\bar
p$ annihilation at rest, as reviewed, e.g., in
\cite{amsler91,amsler98}; in particular, the Crystal Barrel experiment
at LEAR obtained the result \cite{amsler92a} 
\beq
BR(\bar p p \to 2\pi^0) = (6.93 \pm 0.43) \times 10^{-4}
\label{br_pbp_to_2pi0_lear}
\eeq
for $\bar p$ annihilation in liquid hydrogen.  From isospin
invariance, this value would also hold for the hypothetical
annihilation of an $\bar n$ on a free neutron to yield a $2\pi^0$
final state.  Since there is no phase-space suppression of the $\bar p
p \to 2\pi^0$ reaction, a remark on the small branching ratio
(\ref{br_pbp_to_2pi0_lear}) is in order.  The $|2\pi^0\rangle$ state
has a wave function of the form $|2\pi^0\rangle = \chi_I \chi_L$,
where $I$ and $L$ denote the isospin and relative orbital angular
momentum of the pion pair, respectively.  This wave function must be
symmetric under exchange of identical bosons. Since the isospin
Clebsch-Gordon coefficient $\langle I_a I_b I_{a3} I_{b3}|I I_3\rangle
= \langle 1 1 0 0 | 1 0\rangle = 0$, it follows that $|2\pi^0\rangle$
has $I=0$ or $I=2$, both of which are even, so $\chi_I$ is symmetric.
Consequently, $\chi_L$ must also be symmetric, and hence $L$ must be
even.  Therefore, this $|2\pi^0\rangle$ state has $J^{PC} = J^{++}$
with total angular momentum $J=L={\rm even}$.  An $|\bar N N\rangle$
state, where $N=p$ or $N=n$, with nearly minimal center-of-mass energy
$\sqrt{s} \simeq 2m_N$ (e.g., a $|\bar p p\rangle$ state resulting
from a stopping antiproton beam incident on a hydrogen target)
preferentially has $L=0$, and hence $P=-(-1)^L=-1$. Thus, there is a
mismatch between the parity of the dominant, ground-state component in
the initial $|\bar N N\rangle$ state and the parity of the
$|2\pi^0\rangle$ final state. The $\bar N N \to 2\pi^0$ reaction can
proceed, but from an initial $|\bar N N\rangle$ state with $S=1$ and a
kinematically dispreferred $L=1$, coupled to $J=0$ (or $J=2$). This
parity mismatch and resultant suppression contributes to the small
value of the branching ratio in (\ref{br_pbp_to_2pi0_lear}).

Our application of these results is for $\bar n$ annihilation in an
oxygen nucleus in the water of the SK detector, and for this case, one
must take into account the fact that the hadronic products of the
annihilation reaction undergo reactions and absorption while
propagating through the interior of the ${}^{16}$O nucleus. This has
the effect of increasing the branching ratios for two-pion channels
relative to the branching ratios for higher-multplicity pion channels.
A Monte Carlo study of the effect of this intranuclear propagation on
the branching ratios for various hadronic products of $\bar n n$
annihilation was carried out by the SK experiment with the resultant
estimate, for $\bar n n$ annihilation in ${}^{16}$O \ \cite{sk_nnbar}:
\beq
BR(\bar n n \to 2\pi^0)_{{}^{16}{\rm O}} = 1.5 \times 10^{-2} \ .
\label{br_nn_to_2pi0_oxygen}
\eeq
Since the ratios of two-body phase space factors
$R^{(e^+e^-)}_2/R^{(2\pi^0)}_2$ and
$R^{(\mu^+\mu^-)}_2/R^{(2\pi^0)}_2$ are nearly equal (with both being
quite close to unity), our results can also be applied to the ratio
$R^{(\bar n n)}_{\mu^+\mu^-/2\pi^0}$.  Substituting the various inputs
into the right-hand side of Eq. (\ref{nbn_ee_over_nbn_2pi0}), we obtain the
result
\beq
\frac{\Gamma_{\bar n n \to \ell^+\ell^-}}{\Gamma_{\bar n n \to 2\pi^0; {}^{16}{\rm O} } } \simeq 2 \times 10^{-5}
\quad {\rm for} \ \ell=e, \ \mu \ .
\label{nbn_to_ellell_over_nbn_to_2pi0_exp}
\eeq

We next use this experimentally derived ratio for $\Delta B=0$ $\bar n
n$ annihilation processes to obtain an estimate of the ratio of
$\Delta B=-2$ processes $\Gamma_{n n \to \ell^+\ell^-}/\Gamma_{n n \to
  2\pi^0; {}^{16}{\rm O} }$.  The underlying $n - \bar n$ transition
matrix element factor $(\delta m)^2$ divides out in this ratio.  The
further analysis thus involves a study of the degree of overlap
between the $|\bar n n\rangle$ state immediately following the $n -
\bar n$ transition (or equivalently, the state $|\bar n n\rangle$
resulting from the combination of two $|n\rangle_{\rm phys.}$ states)
and the two final states. Since the annihilation occurs on the length
scale of $\sim 1$ fm, a reasonable approximation is to consider the
initial $|nn\rangle$ and $|\bar n n\rangle$ states by themselves,
independent of the other nucleons on the nucleus.  The wave function
of the $|nn\rangle$ state has the form $|nn\rangle = \phi_I \phi_S
\phi_L$, where $I$, $S$, and $L$ denote the isospin, spin, and orbital
angular momentum of the $nn$ di-neutron. This wave function must be
antisymmetric under interchange of identical fermions, so since $I=1$
(symmetric), it follows that the product $\phi_S \phi_L$ must be
antisymmetric under this interchange. The energetically preferred
configuration is the one with lowest energy, i.e., the ground state,
which has $L=0$, so $\phi_L$ is symmetric, and therefore the neutron
spins must combine antisymmetrically to produce $S=0$.  The six-quark
operator in the effective Lagrangian that mediates the $n-\bar n$
transition is a Lorentz scalar and hence does not change $S$ or $L$,
so the $|\bar n n\rangle$ state immediately after this transition also
has $S=L=0$ and hence, in standard spectroscopic notation, is a
${}^1S_0$ state. For a fermion-antifermion pair, $P=-(-1)^L$ and
$C=(-1)^{L+S}$, so this $|\bar n n\rangle$ state has $J^{PC}=0^{-+}$.
This cannot couple directly to the photon, which has $J^{PC}=1^{--}$,
so there is a mismatch in both $J$ and $C$.  The requisite $J^{PC}$
can occur as the result of a spin flip (SF) from $S=0$ to $S=1$. We
incorporate the probability for this in a factor $P_{SF}$.  As
discussed above, for the $|nn\rangle \to |\bar n n\rangle \to
|2\pi^0\rangle$ transition, we use the SK Monte Carlo results.

We thus obtain the improved estimate 
\beq
\Gamma_{nn \to \ell^+\ell^-} = (2 \times 10^{-5}) \, P_{SF} \, \Gamma_{nn \to 2\pi^0}
\quad {\rm for} \ \ell=e, \ \mu \ .
\label{gamma_nn_to_ellell_over_nn_to_2pi0_exp}
\eeq
Combining our result
(\ref{gamma_nn_to_ellell_over_nn_to_2pi0_exp}) with the experimental lower
limit on $\Gamma^{-1}_{nn \to 2\pi^0}$ in
Eq. (\ref{tau_nn_to_2pi0_sk}), we infer the lower bound
\beqs
\Gamma^{-1}_{nn \to \ell^+\ell^-} &\gsim& (2 \times 10^{37}) \, P_{SF}^{-1} \ \ {\rm yrs} \cr\cr
 &>& 2 \times 10^{37} \ {\rm yrs} \quad {\rm for} \ \ell=e, \ \mu \ ,
\label{tau_limit_nn_to_ellell}
\eeqs
where the second line in the inequality (\ref{tau_limit_nn_to_ellell}) is a conservative limit
that just uses the fact that the spin-flip probability $P_{SF} < 1$. 
As was true of the bounds derived in \cite{dnd} and even more so here,
this is much stronger than the direct lower bounds on the partial
lifetimes (from the SK experiment)
\cite{sk_dinucleon_to_ellell}:
\beq
\Gamma^{-1}_{nn \to e^+ e^-} > 4.2 \times 10^{33} \ {\rm yr}
\label{tau_limit_nn_to_ee_sk}
\eeq
and
\beq
\Gamma^{-1}_{nn \to \mu^+ \mu^-} > 4.4 \times 10^{33} \ {\rm yr} \ .
\label{tau_limit_nn_to_mumu_sk}
\eeq
%


\section{Conclusions}
\label{conclusion_section} 

In this paper, using experimental data on $\bar p p$ and $e^+e^-$ annihilation,
we have obtained strengthened lower bounds on the partial
lifetimes for the dinucleon decays $nn \to e^+e^-$ and $nn \to \mu^+\mu^-$. Our
bounds improve upon those in Ref. \cite{dnd} and are considerably stronger than
direct experimental lower bounds on these decays. 


\begin{acknowledgments}

The research of R.S. was supported in part by 
the NSF Grant NSF-PHY-1915093.  R.S. thanks S. Girmohanta for
valuable discussions.

\end{acknowledgments}



\end{document}